\documentclass[aps,pra,showpacs,twocolumn]{revtex4}
\usepackage{}
\usepackage{amsfonts}
\usepackage{amssymb}
\usepackage{color}
\usepackage{amsmath}
\usepackage{graphicx}
\usepackage{subfigure}
\usepackage{txfonts}
\usepackage{color}
\usepackage{ulem}
\usepackage{verbatim}
\usepackage{appendix}

\newcommand{\opa}{{\hat a}}

\begin{document}

\title{Non-Markovianity in the collision model with environmental block}
\author{Jiasen Jin and Chang-shui Yu}
\affiliation{School of Physics, Dalian University of Technology, 116024 Dalian, China}
\begin{abstract}
We present an extended collision model to simulate the dynamics of an open quantum system. In our model, the unit to represent the environment is, instead of a single particle, a block which consists of a number of environment particles. The introduced blocks enable us to study the effects of different strategies of system-environment interactions and states of the blocks on the non-Markovianities. We demonstrate our idea in the Gaussian channels of an all-optical system and derive a necessary and sufficient condition of non-Markovianity for such channels. Moreover, we show the equivalence of our criterion to the non-Markovian quantum jump in the simulation of the pure damping process of a single-mode field. We also show that the non-Markovianity of the channel working in the strategy that the system collides with environmental particles in each block in a certain order will be affected by the size of the block and the embedded entanglement and the effects of heating and squeezing the vacuum environmental state will quantitatively enhance the non-Markovianity.
\end{abstract}

\maketitle

\section{Introduction}
The interaction between a quantum system and an environment leads to a non-unitary time-evolution of the state of system. Such irreversible dynamics can be described by the theory of open systems \cite{breuer2002}. Understanding the dynamics of an open quantum system is an essential question in quantum information processing \cite{verstraete2009,vasile2011a,vasile2011b,matsuzaki2011,chin2012}, quantum biology \cite{ishizaki2005,chin2010} and quantum optics \cite{hoeppe2012,zippilli2014,zhang2016}. Usually the dynamics can be classified as Markovian or non-Markovian cases. In the Markovian case, the dynamics is typically characterized by the master equation in the so-called Lindblad form which corresponding to a completely positive and trace-preserving (CPT) map. In particular, if the Lindbladian of the master equation is time-independent, it gives rise to a dynamical semigroup of maps \cite{breuer2002}. However the Markovian demonstration is not always adequate. For instance, the CPT condition is violated if a quantum system is strongly coupled to the environment, so that the dynamics becomes non-Markovian \cite{devega2017}. In recent years a number of criteria characterizing non-Markovianity have been proposed, from different perspectives, basing on the dynamical divisibility \cite{rivas2010,hou2011,hou2015,torre2015}, back-flow of information characterized by trace distance \cite{breuer2009,vasile2011b}, Fisher information \cite{lu2010}, mutual information \cite{luo2012}, relative entropy \cite{chrusciski2012,chrusciski2014}, accessible information \cite{fanchini2014}, Gaussian interferometric power \cite{souza2015} and response functions \cite{strasberg2017a}. For recent reviews, see \cite{rivas2014,breuer2016}. These criteria or measures help us to distinguish whether a dynamics is Markovian or not; however, in general they do not agree with each other in detecting the emergence and quantifying the degree of non-Markovianity.

An alternative approach to studying the dynamics of an open system is the so-called collision model (CM). In the CM based scheme, the continuous time-evolution of a system is simulated by a sequence of system-environment collisions representing the interactions between system and the environment. The environment is represented by an ensemble of uncorrelated identical particles. If the system collides with each environmental particle in a sequential way, the dynamics of the system is Markovian since the environmental particle in the upcoming collision is fresh and thus contains no information of the history. By introducing the intra-collision between environmental particles \cite{ciccarello2013a,ciccarello2013b,mccloskey2014,jin2015,wang2017,bernardes2017}, long-range system-environment collisions \cite{cakmak2017},  the correlations among environmental particles \cite{rybar2012,bernardes2014,mascarenhas2017}, and a composite structure of the system \cite{lorenzo2017}, the non-colliding environmental particle could have the possibility to carry the information of the history, thus the dynamics of the system may become non-Markovian. Very recently, the idea of CM is also adopted in the content of thermodynamics \cite{strasberg2017b}. We note that the unit representing the environment is a single particle in the aforementioned modified CMs. However, this is abridged in simulating the details of system-environment interactions as well as the diverse states of the environment: on the one hand, the approaches of the memory recover, such as recovering from the latest to the earliest time or the reverse, cannot be manifested through a collision between the system and the single-particle environmental unit; on the other hand, the single-particle unit excludes the nonlocal many-body correlations of the environment. Therefore, a CM with more complicated environmental unit would be of interest not only in simulating the dynamics of open system but also in exploring the essence of the non-Markovian process.

In this work, we will consider an extended CM that the environment is represented by an ensemble of identical blocks. Each block consists of a number of particles. The system-environmental interactions are simulated by the collisions between the system and the environmental blocks. The internal structures of the environmental blocks enable us to explore how the system-environment interactions and the environmental states affect the non-Markovianity of the dynamics. Here, we will mainly discuss our CM in the realization of all-optical system. We will derive a necessary and sufficient condition of the non-Markovianity in our CM basing on the indivisibility of the dynamical maps and show the evidence of its equivalence to the non-Markovian quantum jump through the pure damping process of a single-mode field. Thanks to the internal structure of the environments introduced by the environmental blocks, we will investigate the effects the strategies of the system-environment collisions which are related to the approaches of memory recover in a realistic dynamical process. We will also study how the properties of the environmental state, such as temperature, squeezing and entanglement, affect the non-Markovianity. Because the collisions between different modes are taken place at the beam-splitters (BSs), we can use the Hamiltonian of two-mode linear mixing, $\hat{H}\propto\opa^\dagger\hat{b} + \opa\hat{b}^\dagger$ ($\opa$ and $\hat{b}$ denote different modes), to create such interactions. The usage of the two-mode mixing Hamiltonian in describing the interactions between a bosonic mode and  a reservoir implies the potential of our CM in simulating the dynamics of the open quantum optical system.

This paper is organized as follows. In Sec. \ref{sec2}, we introduce the idea of our model and apply this idea to the Gaussian channel in an all-optical system. In Sec. \ref{sec3A}, we review the measure of non-Markovianity of the Gaussian channel  recently proposed in Ref. \cite{torre2015} and derive an explicit expression of the measure of non-Markovianity for our model. In Sec. \ref{sec3B}, we discuss the non-Markovianity of the Gaussian channel with vacuum environmental state through two strategies of system-environment collisions and simulate the pure damping process of a single-mode field. The necessary and sufficient condition of the non-Markovian channel in the vacuum environment is given as well. In Sec. \ref{sec3C}, we investigate the effects of the temperature and squeezing on the non-Markovianity of the channels with generic Gaussian state. We compare the non-Markovianities of the channels with product and entangled states in in Sec. \ref{sec3D}. The conclusion is drawn in the last section.

\section{Simulating collision model in all-optical system}\label{sec2}
\subsection{The CM with environmental block}

\begin{figure}
  \includegraphics[width=1\linewidth]{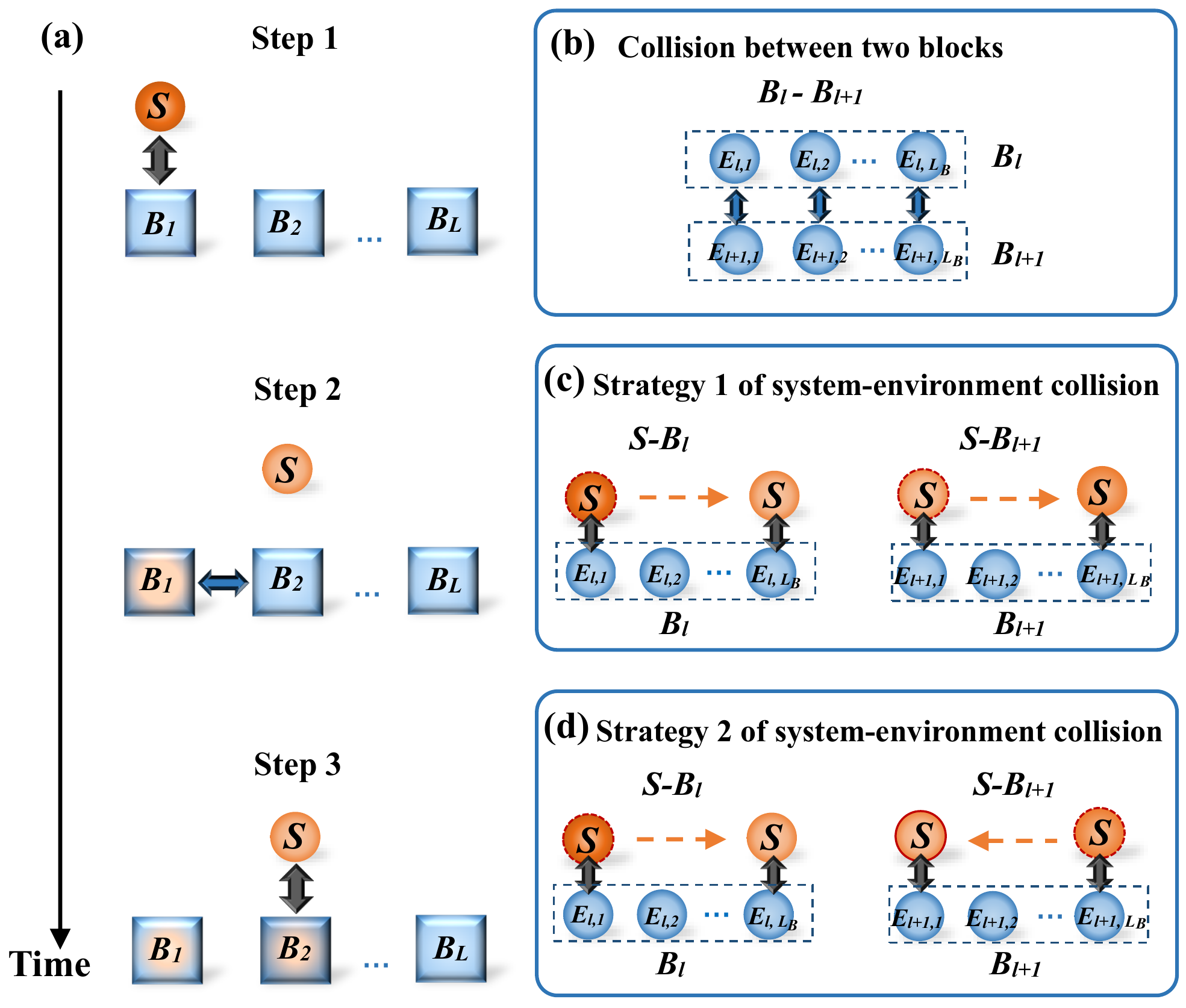}
  \caption{Schematic illustration of the modified collision model. (a) The discrete dynamics of an open system is simulated by a series of collisions  between system and environmental blocks. In step 1, the system particle (circle labeled by ``$S$") collides with block $B_l$ (squares labeled by ``$B$"). The $S$-$B_l$ collision is implemented by collides ``$S$" with each particle $E_{l,j}$ (circles labeled by ``$E$") in $B_l$ in ascending order of $j$. In step 2, block $B_l$ collides with $B_{l+1}$. The $B_l$-$B_{l+1}$ collision is implemented by colliding each pair of $E_{l,j}$ and $E_{l+1,j}$ particles, see (b). In step 3, the system collides with block $B_{l+1}$ and then go to step 2 with $l \rightarrow l+1$ for iteration. There are two strategies of the $S$-$B_{l+1}$ collision. In strategy 1, the $S$-$E_{l+1,j}$ collisions take place in the same order of $j$ as that in $S$-$B_{l}$ collision, i.e. the information firstly input to the environment firstly outputs, see (c); in strategy 2, the sequence of $S$-$E_{l+1,j}$ collisions take place in the order of $j$ which is opposite to that in the $S$-$B_l$ collision, i.e., the information firstly input to the environment lastly outputs, see (d). In Boxes (b)-(d), the system circles with dashed border denote the initial position and the ones with solid border denote the final position. The blue circles $E_{l,j}$ in the dashed rectangle belong to block $B_l$.}
  \label{sketch}
\end{figure}

In our model the unit to represent the environment is a block rather than a single particle as that in the standard CM. An environmental block is consisted of a number of particles and all the blocks are supposed to be identical. For the explicitness of explanation, we label the $l$-th block as $B_l$ with $l=1,2,...,L$ and the particle in $B_l$ as $E_{l,j}$ with $j = 1,2,...,L_B$. We have set the number of blocks to be $L$ and the number of particles in a block to be $L_B$. We will denote $L_B$ to the size of the block hereinafter. As shown in Fig. \ref{sketch}, our model works via the following steps.

{\it Step 1.} As the start, the system collides with the environmental block $B_l$ with $l=1$. The $S$-$B_l$ collision is accomplished by the system sequentially colliding with the particles $E_{l,j}$ in a certain order, for example, the ascending order of $j$. After the $S$-$B_l$ collision, each particle in $B_l$ carries part of the information of system at discrete time points during the past. More precisely, the earliest information of system is stored in $E_{l,1}$ while the latest information is stored in $E_{l,L_B}$.

{\it Step 2.} In this step the intra-collision of the environment takes place. The block $B_l$ collides with a fresh block $B_{l+1}$. The $B_l$-$B_{l+1}$ collision is accomplished by respectively colliding each pair of $E_{l,j}$ and $E_{l+1,j}$ particles. After the $B_l$-$B_{l+1}$ collision, part of the lost information has possibility to be transferred to the block $B_{l+1}$.

{\it Step 3.} The collision between the system $S$ and the block $B_{l+1}$ takes place. The $S$-$B_{l+1}$ collision enables the system to get the lost information back. In this step, we may implement the $S$-$B_{l+1}$ collision in two different strategies. In strategy 1, the system $S$ always sequentially collides with $E_{l+1,j}$s in the same order of $j$ as in step 1, i.e., the information first input to block $B_l$ is first output. In strategy 2, the system collides with $E_{l+1,j}$s in the reverse order of $j$ as that in $S$-$B_l$ collision, i.e. the information first input to the environmental block is last output. Once the $S$-$B_{l+1}$ collision is completed, we go to step 2 with $l\rightarrow l+1$ to iterate.

We would like to emphasize that in our CM the system-environment collision is represented by the $S$-$B_l$ collision. This implies a coarse-graining of the system evolution. We are only interested in the evolved system state after each complete $S$-$B_l$ collision, and the intermediate system states after each $S$-$E_{l,j}$ collisions are considered to be the details hiding in the system-environment collision.

\subsection{A scheme in the all-optical system}
Our CM can be implemented in the all-optical system which is composed of an array of BSs. The realistic optical system can be perfectly controlled, integrated and scaled up. The system and environmental particles are represented by the independent optical modes propagating along different paths. The collisions between any two particles can be realized by mixing two corresponding input modes at the BS. We recall that a BS transfers two input modes ${\hat {\bf a}}^{\mathrm{ in}} = [\opa_1^{\mathrm{ in}},\opa^{\mathrm{ in}}_2]^{\text{T}}$ into output modes ${\hat {\bf a}}^{\mathrm{ out}}=[\opa_1^{\mathrm{ out}},\opa_2^{\mathrm{ out}}]^{\text{T}}={\cal S}{\hat {\bf a}}^{\mathrm{ in}}$, where $\opa$ is the annihilation operator (the superscript T denotes the transpose), and ${\cal S}$ is the scattering matrix
\begin{equation}
{\cal S}=\left[
          \begin{array}{cc}
            r & t \\
            -t & r \\
          \end{array}
        \right],\label{BS}
\end{equation}
with $r=\sin{\theta}$ and $t=\cos{\theta}$ being the reflectivity and transmissivity of the BS and $\theta\in[0,\pi/2]$. In Eq. (\ref{BS}), we have set the reflected mode to be the output mode. For $\theta = \pi/2$, both the input modes will be completely reflected thus indicating the strength of interaction between two modes is zero, while for $\theta = 0$, both the input modes will completely transmit thus indicating a swap operation of two modes.

In our model we denote the BS that mixes the system and environmental modes as BS1 with the reflectivity and transmissivity being $r_1=\sin{\theta_1}$ and $t_1=\cos{\theta_1}$, and the BS that mixes two environmental modes as BS2 with the reflectivity and transmissivity being $r_2=\sin{\theta_2}$ and $t_2=\cos{\theta_2}$. Therefore, the strengths of system-environment and environment-environment interactions can be tuned by varying $\theta_1$ and $\theta_2$, respectively. In the following, we will describe how to realize the $S$-$B_l$, $B_l$-$B_{l+1}$ and $S$-$B_{l+1}$ collisions in the all-optical system.

{\it $S$-$B_l$ collision}. The collision between the system $S$ and the environmental block $B_l$ can be simulated by sequentially mixing the system mode, $\opa_S$, and each environmental mode, $\opa_{l,j}$, at a series of identical BS1s. For instance, in the case of $S$ interacts with each $E_{l,j}$ in the ascending order of $j$, the $\opa_S$ mode firstly interacts with $\opa_{l,1}$ at the first BS1s, and then the output (reflected mode) of $\opa_S$ will interact with $\opa_{l,2}$ at the second BS1 and so on so forth. The $S$-$B_l$ collision is completed until the $\opa_S$ mode has interacted with the $\opa_{l,L_B}$ mode.

{\it $B_l$-$B_{l+1}$ collision}. In order to simulate the $B_l$-$B_{l+1}$ collision, each of the output (reflected) mode of $\opa_{l,j}$ is guided to interact with the corresponding $\opa_{l+1,j}$ mode of $B_{l+1}$ individually. It should be noted that, different from the $S$-${B_l}$ collision, the interactions between $\opa_{l,j}$ and $\opa_{l+1,j}$ take place at BS2s.

{\it $S$-$B_{l+1}$ collision}. As mentioned before, there are two strategies for the $S$-$B_{l+1}$ collisions. For strategies 1 and 2, the output mode of $\opa_S$ is guided to interact with the output of $\opa_{l+1,j}$ modes in the same and reverse orders of $j$ to that in the previous $S$-$B_l$ collision, respectively. Again the $S$-$B_{l+1}$ collisions take place at BS1s.

\begin{figure}
  \includegraphics[width=.95\linewidth]{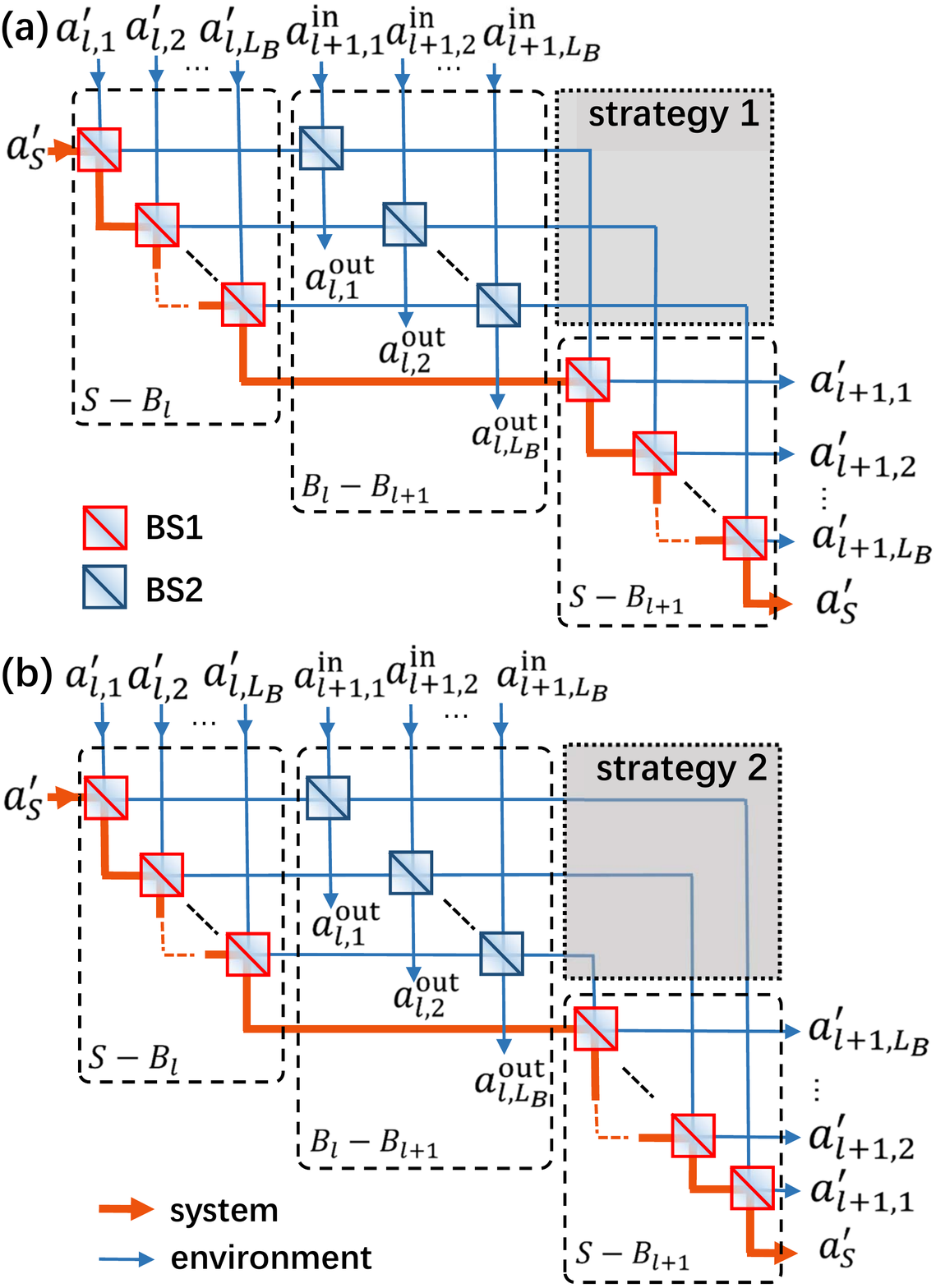}
  \caption{Schematics of the setups working in strategy 1 (a) and in strategy 2 (b). The red and blue squares denote BS1 and BS2, respectively. The solid lines denote the system mode in red and the environmental modes in blue. The operators with primes denote the intermediate evolved states and the operators with superscripts `in' and `out' denote the initial and final states. In both (a) and (b), the dashed boxes are the building blocks simulating the $S$-$B_l$, $B_l$-$B_{l+1}$ and $S$-$B_{l+1}$ collisions. In the $S$-$B_l$ collision, the system mode sequentially interacts with the ($l,j$)-th environmental modes at BS1s. After the $S$-$B_l$ collision, the output of the ($l,j$)-th environmental modes, i.e. the reflected modes, are guided to interact with the fresh ($l+1,j$)-th modes of $B_{l+1}$ individually to simulate the $B_l$-$B_{l+1}$ collisions. In the $S$-$B_{l+1}$ collision, the output of ($l,j$)-th modes are discarded and the output of ($l+1,j$)-th modes are guided to interact with the system mode. The difference between strategies 1 and 2 are presented in the dotted boxes (shaded gray). In strategy 1, the interactions between system and the ($l+1,j$)-th modes are implemented with the same order of $j$ to that in the previous $S$-$B_l$ collision, while in strategy 2, the interactions between system and the ($l+1,j$)-th modes are implemented with the reverse order of $j$ to that in the previous $S$-$B_l$ collision. By concatenating the $B_l$-$B_{l+1}$ and $S$-$B_{l+1}$ building blocks, the stroboscopic evolution of the system mode can be simulated.}
  \label{illustration1}
\end{figure}

In Fig. \ref{illustration1}, we show the setups working in strategies 1 and 2, respectively. Both setups are composed of the building blocks (dashed boxes) that realize the $S$-$B_l$, $B_{l}$-$B_{l+1}$ and $S$-$B_{l+1}$ collisions. The operators with primes denote the intermediate evolved state and the operators with superscripts `in' and `out' denote the initial and final state, respectively. The dotted boxes (shaded gray) show the distinction of strategies 1 and 2, i.e. the system mode interacts with environment modes in different orders of $j$.

We can concatenate such processes to simulate the discrete dynamical evolution of the system mode. Suppose the system will interact with $L$ environmental blocks, then the channel can be described, with the help of scattering matrix ${\cal S}(L)$, by $[\opa_S^{\text{out}},\opa_{1,1}^{\text{out}},...,\opa_{l,j}^{\text{out}},...,\opa_{L,L_B}^{\text{out}}]^{\text{T}}={\cal S}(L)[\opa_S^{\text{in}},\opa_{1,1}^{\text{in}},...,\opa_{l,j}^{\text{in}},...,\opa_{L,L_B}^{\text{in}}]^{\text{T}}$.
Hereinafter, we use the subscript $S$ to denote the system mode and the pairwise $(l,j)$ to denote the $j$-th mode in the $l$-th environmental block.

The $(L_BL+1)$-dimensional scattering matrix ${\cal S}(L)$, for $L\ge2$,has the form
\begin{equation}
{\cal S}(L) = {\cal S}_{SB_L}\left(\prod_{l=1}^{L-1}{\cal S}_{B_lB_{l+1}}{\tilde{\cal S}}_{SB_l}\right),\label{ScatteringMatrix}
\end{equation}
where the matrices in the cumulative product are in the descending order of $l$ from right to left. In Eq. (\ref{ScatteringMatrix}), the matrix ${\cal S}_{B_lB_{l+1}}$ describing $B_l$-$B_{l+1}$ collision is given by
\begin{equation}
{\cal S}_{B_lB_{l+1}} = \left[
      \begin{array}{cccc}
        I_{(l-1)L_B+1} & 0 & 0 & 0 \\
        0 & r_2I_{L_B} & t_2I_{L_B} & 0 \\
        0 & -t_2I_{L_B}  & r_2I_{L_B} & 0 \\
        0 & 0 & 0 & I_{(L-l-1)L_B} \\
      \end{array}
    \right],\label{ScatteringMatrixBlBl1}
\end{equation}
where $I_n$ is the $n\times n$ identity matrix. The matrix $\tilde{{\cal S}}_{SB_l}$ describing the $S$-$B_l$ collision has two different forms with respect to strategies 1 and 2. In strategy 1, we have $\tilde{{\cal S}}_{SB_l} = {\cal S}_{SB_l}$ for all $l$ with ${\cal S}_{SB_l}=\prod_{j=1}^{L_B}{{\cal S}_{l,j}}$, while in strategy 2, we have $\tilde{{\cal S}}_{SB_l} = {\cal S}_{SB_l}$ for odd $l$ and $\tilde{{\cal S}}_{SB_l} = {\cal R}_{SB_l}$ for even $l$ with ${\cal R}_{SB_l}=\prod_{j=L_B}^{1}{{\cal S}_{l,j}}$. The matrix ${\cal S}_{l,j}$ describing the interaction between system and the ${(l,j)}$-th environmental mode is given by
\begin{equation}
{\cal S}_{l,j} = \left[
      \begin{array}{cccc}
        r_1 & 0 & t_1 & 0 \\
        0 & I_{(l-1)L_B+(j-1)} & 0 & 0 \\
        -t_1 & 0 & r_1 & 0\\
        0 & 0 & 0 & I_{(L-l)L_B+(L_B-j)}\\
      \end{array}
    \right]
    \label{Slj}.
\end{equation}
Note that ${\cal S}_{SB_l}$ and ${\cal R}_{SB_l}$ are obtained by multiplying ${\cal S}_{l,j}$ in the ascending and descending orders of $j$, respectively. From Eqs. (\ref{ScatteringMatrix})-(\ref{Slj}), we see that the property of the `bare' channel (i.e. with vacuum environment) is determined by the reflectivities and transmissivities of BS1 and BS2.

\subsection{The dynamics of the system mode}
It is convenient to study the dynamics in the channel with the characteristic function formalism \cite{walls1994}. Actually, the density operator ${\hat \rho}$ of a quantum state is equivalent to the characteristic function in presenting the probability distribution. The symmetrically ordered characteristic function is defined by $\chi(\nu)=\text{Tr}[{\hat D}(\nu) {\hat \rho}]$ with the Weyl displacement operator ${\hat D}(\nu)=\exp{(\nu\opa^\dagger-\nu^*\opa)}$. Thus we can represent the density operator of a bosonic system which is defined in an infinite-dimensional Hilbert space with a complex function. In particular, in terms of the characteristic function, the first and second moments are sufficient to characterize the Gaussian state which is widely used in the quantum information processing with continuous variables system \cite{ferraro2005}. Reversely, the density operator $\hat{\rho}$ can be represented in the Weyl expansion with the $\chi(\nu)$ acting as the weight function, i.e., $\hat{\rho}=\int{\text{d}^2\nu\chi(\nu)\hat{D}(\nu)/\pi}$. In our model, the joint characteristic function of the multimode input state ${\hat \rho}_{J}^{\mathrm{in}}$ is given by $\chi_{J}^{\mathrm{in}}(\vec{\nu})=\text{Tr}[{\hat D}_{J}({\vec \nu}){\hat \rho}_{J}^{\mathrm{in}}]$ with $\vec{\nu} = [\nu_S,\nu_{1,1},...,\nu_{L,L_B}]^{\text{T}}$ being a complex vector and ${\hat D}_{J}(\vec{\nu})={\hat D}_S(\nu_S)\bigotimes_{l=1}^{L}\bigotimes_{j=1}^{L_B}{\hat D}_{l,j}(\nu_{l,j})$. The subscript `$J$' denotes the joint modes.

Initially the modes of the system and the environment are uncorrelated, the joint input characteristic function is thus calculated by
\begin{equation}
\chi^{\mathrm{in}}_{J}(\vec{\nu})=\chi^{\mathrm{in}}_S(\nu_s)\times\prod_{l=1}^{L}{\chi^{\mathrm{in}}_{l}(\vec{\nu}_{l})},
\label{chiin}
\end{equation}
where $\vec{\nu}_l = [\nu_{l,1},\nu_{l,2},...,\nu_{l,L_B}]^{\text{T}}$ and $\chi^{\mathrm{in}}_l(\vec{\nu}_l)$ is the characteristic function of the $l$-th block.

The input-output relation of the joint characteristic function after $L$ times system-environment collisions is just determined by the scattering matrix ${\cal S}(L)$ through the following formula, as detailed in Appendix,
\begin{equation}
\chi^{\mathrm{out},L}_{J}(\vec{\nu})=\chi_{J}^{\mathrm{in}}[{\cal S}^{-1}(L)\vec{\nu}],\label{chiout}
\end{equation}
where $\chi^{\mathrm{out},L}_{J}(\vec{\nu})$ is the joint characteristic function of the output modes and ${\cal S}^{-1}(L) = {\cal S}^{\dagger}(L)$ is the inverse of the scattering matrix ${\cal S}(L)$.
Since we are interested in the evolution of the system mode, we need to trace out all the environmental modes in Eq. (\ref{chiout}).
According to the Theorem 2 of Ref. \cite{wang2007}, the partial trace over all the environmental modes of $\chi^{\mathrm{out},L}_{J}(\vec{\nu})$ can be done by setting $\vec{\nu}=[\nu_S,0,...,0]^{\text{T}}$ . Thus we can obtain the reduced characteristic function of the output system mode $\chi_{S}^{\text{out},L}(\nu_S)$  as
\begin{eqnarray}
\chi_{S}^{\text{out},L}(\nu_S)&=&\chi^{\mathrm{out},L}_{J}(\vec{\nu})\big|_{\vec{\nu}=[\nu_S,0,...,0]^{\text{T}}}\cr\cr
&=&\chi^{\text{in}}_{J}\left[{\cal S}^{-1}(L)\vec{\nu}\right]\big|_{\vec{\nu}=[\nu_S,0,...,0]^{\text{T}}}\label{chiSout}\cr\cr
&=&\chi^{\text{in}}_{J}(\vec{c}_1\nu_S),
\label{chiSout}
\end{eqnarray}
where $\vec{c}_1 = [c_{1,1}, c_{1,2}, ..., c_{1,L_BL+1}]^{\text{T}}$ is a column vector that equals to the transpose of the first row of ${\cal S}(L)$.

Here we concentrate on the cases that all the input modes are initially in Gaussian states. A state of continuous variable system is Gaussian if its characteristic function is Gaussian. The Gaussian state are the resources for a plethora of quantum information and communication protocols with continuous variables \cite{ferraro2005}. In our model, it is easy to prove that the channel with Gaussian environmental state will always keep the Gaussianity of the system state, therefore we may regard the channel described by Eq. (\ref{ScatteringMatrix}) to be a Gaussian channel. We recall that the characteristic function of a generic Gaussian states is given by
\begin{equation}
\chi^{\mathrm{in}}_{\text{G}}(\nu)=\exp{\left[-\left(A+\frac{1}{2}\right)|\nu|^2-\frac{1}{2}\left(B^*\nu^2+B\nu^{*2}\right)+C\nu^*-C^*\nu\right]}.
\label{chiX}
\end{equation}
The real parameter $A$ and complex parameters $B$ and $C$ are related with the properties of the Gaussian state as
\begin{eqnarray}
A&=&\left(n+\frac{1}{2}\right)\cosh{(2r)}-\frac{1}{2},\cr\cr
B&=&-\left(n+\frac{1}{2}\right)\sinh(2r)e^{i\phi},\cr\cr
C&=&\alpha,
\label{GaussianABC}
\end{eqnarray}
where $n$ is the thermal mean photon number, $r$ is the squeezing strength, $\phi$ is the rotating angle and $\alpha$ is the complex displacement \cite{marian2004}.

So far, we are able to describe the dynamics of the system mode with the help of the scattering matrix in the characteristic function formalism. In our CM, the correlations between the system and each environmental blocks are built after each $S$-$B_l$ collisions and present during the whole evolution of the system mode, because all the environmental modes are traced out after the $S$-$B_L$ collision. This is different to the existing CMs in which the system-environment correlations are erased, before or after the $B_l$-$B_{l+1}$ collision, in each step \cite{mccloskey2014,cakmak2017}. It is shown that the system-environment correlations play important role in establishing the non-Markovianity \cite{mccloskey2014}, thus our CM has the potential in studying the role of system-environment correlations in a rather flexible way.

\section{non-Markovianity of the Gaussian channel}\label{sec3}
\subsection{Measure of non-Markovianity}\label{sec3A}
In Ref. \cite{torre2015}, a measure of non-Markovianity of the Gaussian channel by quantifying the degree of the violation of dynamical divisibility is presented. We will employ this measure in our model. According to Eq. (\ref{chiSout}), we can represent the evolved system mode after $l$ times system-environment collisions with the following dynamical map on the input characteristic function,
\begin{equation}
\chi_S^{\text{in}}(\nu_S)\mapsto\chi^{\text{out},l}_S(\nu_S)={\cal E}_l[\chi_S^{\text{in}}(\nu_S)],
\end{equation}
or, in terms of the covariance matrix,
\begin{equation}
\sigma_S^{\text{in}}\mapsto \sigma_S^{\text{out},l}={\cal E}_l\left[\sigma_S^{\text{in}}\right].
\end{equation}
The covariance matrix is the second moment of the characteristic function and its elements are defined by
\begin{equation}
\sigma_{i,j}:= \frac{1}{2}\langle\{\Delta {\hat x}_i,\Delta {\hat x}_j\}\rangle,
\end{equation}
where $\{\cdot,\cdot\}$ is the anticommutator, $\langle\cdot\rangle$ is the expected value and $\Delta {\hat x}_i = {\hat x}_i-\langle {\hat x}_i\rangle$ with ${\hat x}_1=(\opa_S+\opa_S^\dagger)/\sqrt{2}$ and ${\hat x}_2=(\opa_S-\opa_S^\dagger)/\sqrt{2}i$.  The symmetrically ordered moments can be computed can be computed as
\begin{equation}
\langle(\opa_S^\dagger)^p\opa_S^q\rangle_{\text{symm}}= (-1)^{p+q}\frac{\partial^{p+q}}{\partial\nu_S^p\partial\nu_S^{*q}}\chi_S(\nu_S)\Big|_{\nu_S=0},
\end{equation}
where the subscript ``symm" denotes the symmetrical order.

The dynamical map ${\cal E}_l$ is always CPT and can be always formally split as the following,
\begin{equation}
{\cal E}_{l} = \Phi_{l,l-1}\circ{\cal E}_{l-1},
\end{equation}
where $\Phi_{l,l-1}$ is an intermediate process that maps the $\chi_S^{\text{out},l-1}$ to $\chi_S^{\text{out},l}$, and the ``$\circ$" represents the composition of the maps. The divisibility of the Gaussian channel can be determined by $\Phi_{l,l-1}$. If $\Phi_{l,l-1}$ is CPT for all $l$, then the dynamics is divisible and hence Markovian. Otherwise, if $\Phi_{l,l-1}$ is non-CPT for some values of $l$, then the dynamics is indivisible and hence non-Markovian.

For a generic Gaussian channel, ${\cal E}_l$ has the following form,
\begin{equation}
{\cal E}_l\left[\sigma_S^{\text{in}}\right]=X_l\sigma_S^{\text{in}}X^{\text{T}}_l+Y_l,\label{covmatrix}
\end{equation}
where $X_l$ and $Y_l$ are $2\times2$ real matrices. The necessary and sufficient conditions of the CPT property of $\Phi_{l,l-1}$ is the semi-positive definiteness of the following $2\times2$ matrix \cite{lindblad2000},
\begin{equation}
\Lambda_l=Y_{l,l-1}-\frac{i}{2}\Omega+\frac{i}{2}X_{l,l-1}\Omega X^{\text{T}}_{l,l-1},
\label{matixLambda}
\end{equation}
with $X_{l,l-1}=X_{l}X^{-1}_{l-1}$, $Y_{l,l-1}=Y_{l}-X_{l,l-1}Y_{l-1}X^{\text{T}}_{l,l-1}$, and $\Omega=[0,1;-1,0]$ being the single mode symplectic matrix. The negative eigenvalue of $\Lambda_l$ contributes to the non-CPT of $\Phi_{l,l-1}$ and, as a consequence, the non-Markovianity of the Gaussian channel. Thus the non-Markovianity of the Gaussian channel can be measured by the sum of the negative eigenvalues of all the $\Lambda_l$,
\begin{equation}
{\cal N}(L) = \sum_{l=2}^{L}{\sum_{k=\pm}{\frac{|\lambda_{l,k}|-\lambda_{l,k}}{2}}},\label{measureNM}
\end{equation}
where $\lambda_{l,k}$ are the eigenvalues of $\Lambda_l$.

Eq. (\ref{measureNM}) is the expression of the non-Markovianity measure for our CM and will be used in the analysis hereinafter. We would like to point out that although Eq. (\ref{measureNM}) is sufficient and necessary in characterizing and quantifying the non-Markovianity, it is computable only when the channel can be completely characterized.  Fortunately, it is possible to completely demonstrate the Gaussian channel of our CM in the all-optical system.

We restrict the initial system state to be a Gaussian state with the characteristic function as expressed in Eq. (\ref{chiX}). The corresponding covariance matrix is given by
\begin{equation}
\sigma_S^{\text{in}}=\left[
      \begin{array}{cc}
        A_S+\frac{1}{2}-\text{Re}(B_S) & -\text{Im}(B_S)  \\
        -\text{Im}(B_S) & A_S+\frac{1}{2}+\text{Re}(B_S) \\
      \end{array}
    \right].
\end{equation}
Once the scattering matrix ${\cal S}(l)$ is constructed and the environmental Gaussian state is specified to $A_{l,j}=A_E$, $B_{l,j}=B_E$, and $C_{l,j}=C_E$, we can compute the evolved characteristic function of the system mode with the help of Eq. (\ref{chiout}) and then obtain the corresponding covariance matrix as the following,
\begin{equation}
\sigma_S^{\text{out},l}=\left[
      \begin{array}{cc}
        A_S(l)+\frac{1}{2}-\text{Re}(B_S(l)) & -\text{Im}(B_S(l))  \\
        -\text{Im}(B_S(l)) & A_S(l)+\frac{1}{2}+\text{Re}(B_S(l)) \\
      \end{array}
    \right],
\end{equation}
with $A_S(l)=(A_S+1/2)c_{1,1}^2(l)+A_l$, $B_S(l)=B_Sc_{1,1}^2(l)+B_l$, and $C_S(l)=C_Sc_{1,1}(l)+C_E\sum_{k=1}^{L_BL+1}{c_{1,k}(l)}$. We have set $A_l=(A_E+1/2)(1-c^2_{1,1}(l))$, $B_l=B_E\sum_{k=2}^{L_BL+1}{c^2_{1,k}(l)}$ and $c_{1,k}(l)$ to be the matrix element of ${\cal S}(l)$ at the 1st row and $k$-th column. Accordingly, we have the explicit forms of $X_l$ and $Y_l$ in Eq. (\ref{covmatrix}) as
\begin{equation}
X_l=\left[
      \begin{array}{cc}
        \text{Re}(c_{1,1}(l)) & -\text{Im}(c_{1,1}(l))  \\
        \text{Im}(c_{1,1}(l)) & \text{Re}(c_{1,1}(l)) \\
      \end{array}
    \right],\label{covmatrixX}
\end{equation}
and
\begin{equation}
Y_l=A_lI_2+\left[
      \begin{array}{cc}
        -\text{Re}(B_l) & -\text{Im}\left(B_l\right)  \\
        -\text{Im}\left(B_l\right) & \text{Re}(B_l) \\
      \end{array}
    \right].\label{covmatrixY}
\end{equation}

The matrices $X_l$ and $Y_l$ can fully demonstrate the Gaussian channel. One can see that the properties of the Gaussian channel is determined by the $r_1=\sin{\theta_1}$ and $r_2=\sin{\theta_2}$ in terms of the $c_{1,k}(l)$ as well as the properties of Gaussian environmental state in terms of $A_E$ and $B_E$. It is straightforward to obtain the eigenvalues of $\Lambda_l$ as a function of $A_E$, $B_E$, and $c_{1,1}(l)$,
\begin{equation}
\lambda_{l,\pm}=\frac{1}{2}\left(2A_E+1\pm\sqrt{4|B_E|^2+1}\right)\left[1-\frac{c_{1,1}^2(l)}{c^2_{1,1}(l-1)}\right].
\label{lambda12}
\end{equation}

\subsection{Vacuum environmental state}\label{sec3B}
We start with the vacuum environmental state, i.e. $A_E$ and $B_E$ are both zero. For vanishing $A_E$ and $B_E$, the eigenvalues of $\Lambda_l$, Eq. (\ref{matixLambda}), are $\lambda_{l,+}=1-c^2_{1,1}(l)/c^2_{1,1}(l-1)$ and $\lambda_{l,-} = 0$. Using Eq. (\ref{measureNM}), we can obtain the non-Markovianity of the channel with vacuum environment, ${\cal N}_{\text{vac}}(L)$, as
\begin{equation}
{\cal N}_{\text{vac}}(L)=\sum_{l=2}^{L}{\max{\left[0,1-\frac{c^2_{1,1}(l)}{c^2_{1,1}(l-1)}\right]}}.
\label{N0}
\end{equation}
The above expression indicates a necessary and sufficient condition of the non-Markovian Gaussian channel with vacuum environmental state, i.e.,
\begin{equation}
|c_{1,1}(l)|\ge |c_{1,1}(l-1)|, \forall l\ge 2.
\end{equation}

\begin{figure}
  \includegraphics[width=1\linewidth]{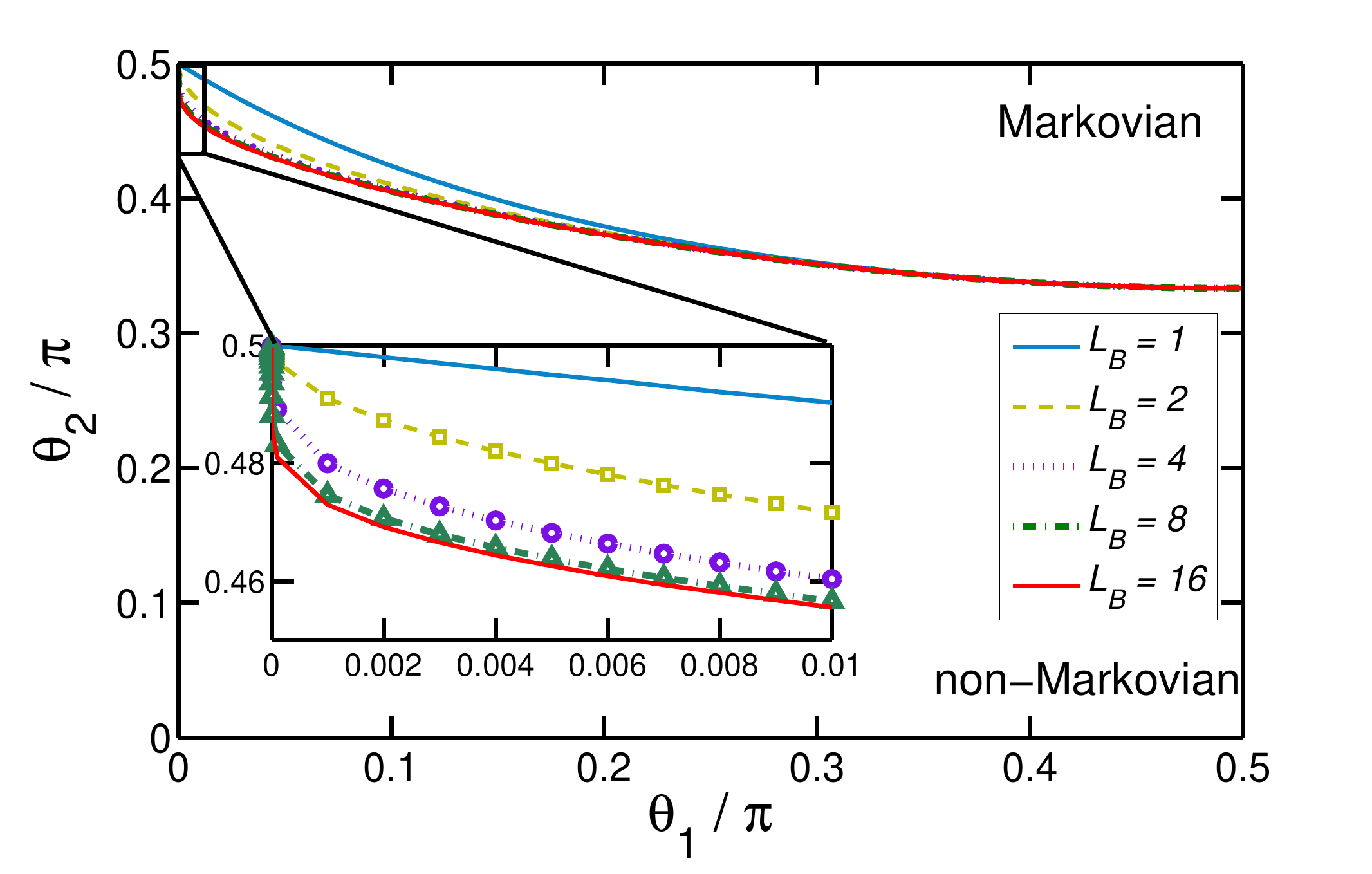}
  \caption{The Markovian and non-Markovianity regions in the plane expanded by $\theta_1$ and $\theta_2$ for different size of environmental block in strategy 1. The environments are vacuum. The stroboscopic evolution is cut off after the system colliding with $L=50$ blocks. For $\theta_1/\pi=0.5$, the system mode is isolated from the environments. For the limit of $\theta_2 = 0$ the evolution of the system mode is unitary and $\theta_2/\pi = 1$ the dynamics of the system is Markovian. With the size of block increasing, the non-Markovian range shrinks and converges for $L_B > 8$. The inset is a zoom in of the range of small $\theta_1/\pi$, i.e., the strong coupling of the system and environment.}
  \label{Nr1r2}
\end{figure}

In Fig. \ref{Nr1r2}, we show the Markovian and non-Markovian regions in the plane expanded by $\theta_1$ and $\theta_2$ in strategy 1 for different sizes of the environmental block. For $L_B = 1$, our model is reduced to the standard CM and thus the values of $\theta_1$ and $\theta_2$ characterize directly the strengths of the system-environment and environment-environment interactions. For the case of $\theta_1/\pi=0.5$, the system mode is complete reflected after each $S$-$B_l$ collision and thus isolated from the environment. As $\theta_1$ decreasing, the system-environment interaction is activated. For the limit case of $\theta_2/\pi=0.5$, the dynamics of system is Markovian since the strength of $B_l$-$B_{l+1}$ collision is zero. In the opposite side, $\theta_2/\pi=0$, the dynamics of system is strongly non-Markovian since the $B_l$-$B_{l+1}$ collision is a perfect swap operation. As a consequence, for a fixed $\theta_1$, we can switch the channel from Markovian to non-Markovian cases by tuning $\theta_2$. There are critical $\theta_2$s that separate the Markovian and non-Markovian regions.

We remind that the CM simulates the dynamics of an open quantum system in a stroboscopic way, i.e., the time interval between two successive system-environment collisions is $\tau$. The overall effect of one collision between system and an environmental block is mapping the system state at time $t$ to $t+\tau$, regardless of the microscopic details in the collision. Namely, we may regard two $S$-$B_l$ collisions with different $L_B$ to be equivalent if they map the same initial state to the same final state. Basing on this idea, we are able to study the cases of $L_B > 1$ in a unified frame. This can be realized in the following approach: if the reflectivity of BS1 is $r_1$ for $L_B = 1$, then the reflectivity of BS1 is set to be $r_1^{1/L_B}$ for $L_B > 1$. This guarantees the identity of the effective strengths (or $\theta_{1,\text{eff}}$) of the system-environment interactions with different $L_B$, because the successive $S$-$E_{l,j}$ collisions in block $B_l$ is Markovian. From Fig. \ref{Nr1r2} we see that the non-Markovian region shrinks with the size of block increasing. However the boundary of non-Markvoian region converges for large $L_B$. In the plot we numerically compute the critical $\theta_2$ as a function of $\theta_1$ with the size of block up to $L_B = 16$. For the case of weak coupling of system and environment the boundary converges fast, while for strong coupling of system and environment the boundary converges slow.

We note that for $L_B = 1$ the channel with strategy 2 is equivalent to that with strategy 1. However, for strategy 2, the non-Markovian region in $\theta_1$-$\theta_2$ plane does not affected by size of the environmental block. We will quantitatively investigate the non-Markovianties in both strategies.

\subsubsection{Non-Markovianities in strategies 1 and 2}
\begin{figure}
  \includegraphics[width=1\linewidth]{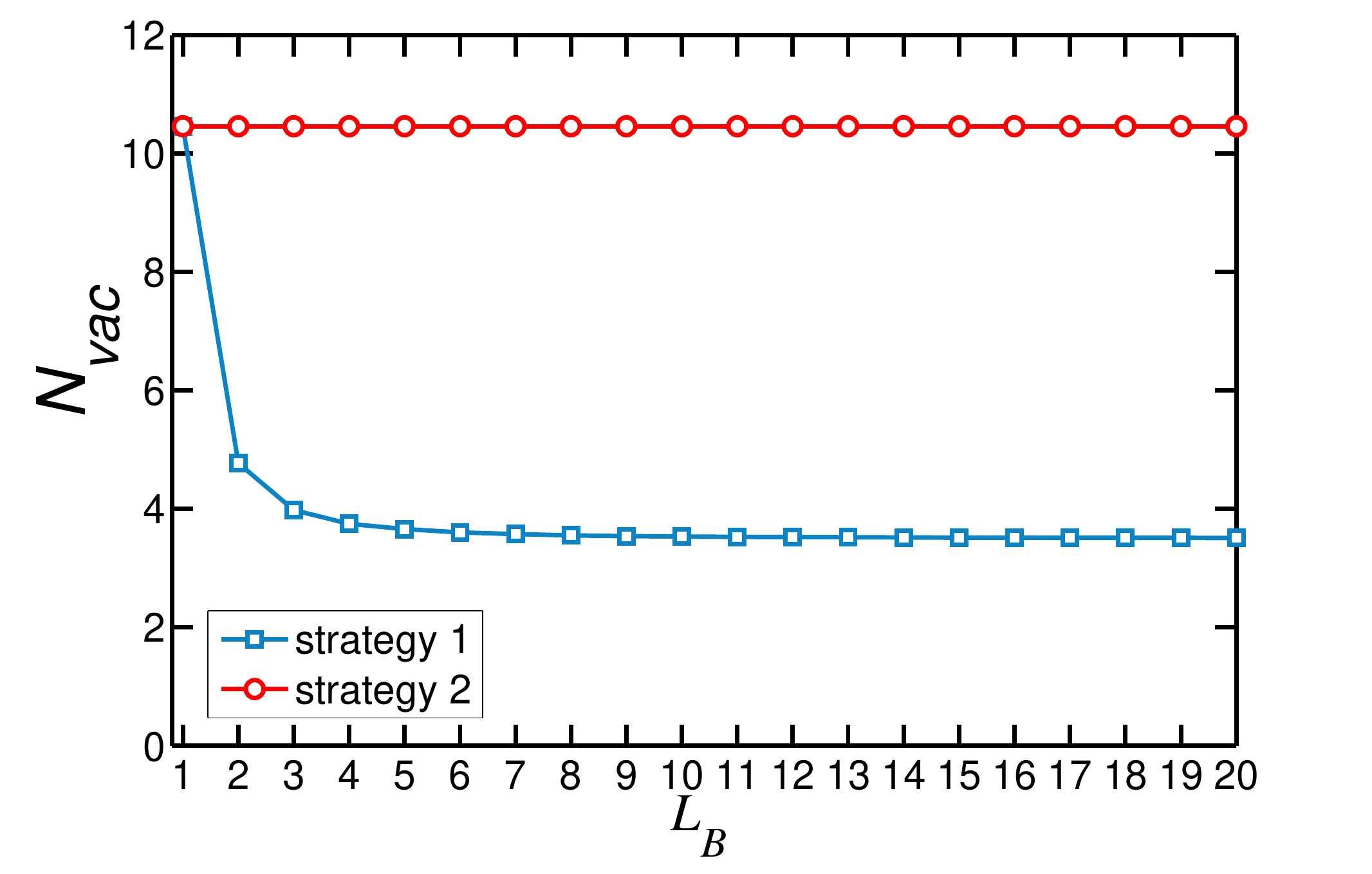}
  \caption{Non-Markovianities of Gaussian channel with vacuum environmental state as a function of $L_B$. The parameters are chosen as $L = 50$,  $\theta_1/\pi = 1/6$ and $\theta_2/\pi = 1/6$.}
  \label{N0NB}
\end{figure}

In this subsection, we will compare the non-Markovianties of both strategies 1 and 2 for, $L_B > 1$, with the vacuum environmental state being vacuum. With the help of Eq. (\ref{N0}), we could compute the non-Markovianities for the channels with both strategies. In Fig. \ref{N0NB} we show the non-Markovianity as a function of $L_B$ with $\theta_1/\pi=\theta_2/\pi=1/6$. The degrees of non-Markovianities of strategy 2 are always stronger than those of strategy 1. Moreover, in strategy 2, the non-Markovianity remains the same as that in the case of $L_B=1$, while, in strategy 1, the non-Markovianity decreases with the size of the block increasing and converges for $L\ge8$. Note that, in strategy 2, the successive $S$-$B_{l-1}$, $B_{l-1}$-$B_l$ and $S$-$B_l$ collisions construct an $L_B$-level nested Mach-Zehnder interferometer of the system mode and $L_B$ (dissipative) environmental modes. Considering the normalization on the reflectivity of $BS1$, i.e. the effective strength of the $S$-$B_l$ interactions for different $L_B$ are equal, we can conclude that the non-Markovianity in strategy 2 is independent of the block size.

\begin{figure}
  \includegraphics[width=1\linewidth]{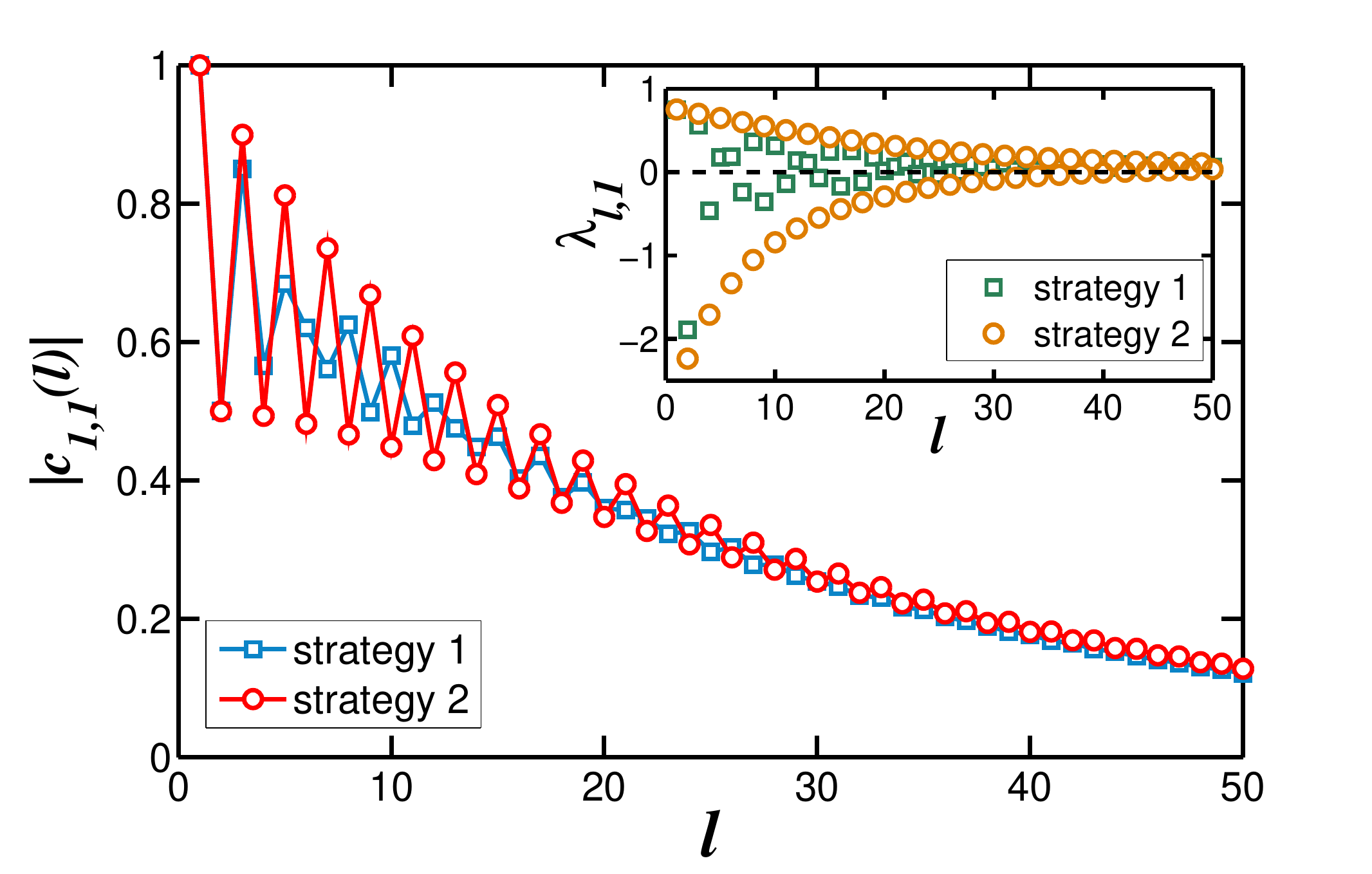}
  \caption{Stroboscopic evolutions of $|c_{1,1}(l)|$ in both strategies for $L_B=16$. The The inset shows the nonzero eigenvalue of $\Lambda_l$ in the stroboscopic evolution. The contribution of the negative eigenvalues in strategy 2 is larger than that in strategy 1. This indicates that the Gaussian channel with strategy 2 violates the divisibility stronger than strategy 1. The parameters of BSs are $\theta_1/\pi = 1/6$ and $\theta_2/\pi = 1/6$. }
  \label{NB10cL}
\end{figure}

In order to show the differences between the two strategies, we show the stroboscopic evolutions of the matrix element $|c_{1,1}(l)|$ and the nonzero eigenvalue of $\Lambda_l$ in Fig. \ref{NB10cL}. We see that in strategy 2 the revival of $|c_{1,1}(l)|$ is stronger than that in strategy 1. As a consequence, the negative eigenvalues contribute more to the indivisibility of the channel in strategy 2. It is easy to understand the advantages of non-Markovianity in strategy 2 in the limit of $\theta_2=0$. In such a case, the time evolution of the system is unitary. Moreover the output of $\opa_{l,j}$ after $S$-$B_l$ collision are the input, with an additional $\pi$ phase, of $\opa_{l+1,j}$ in $S$-$B_{l+1}$ collision. This guarantees the time-reversal symmetry of the input and output of system states in two consecutive system-block collisions for strategy 2 and leads a strong non-Markovianity.

\subsubsection{Pure damping process of a single-mode field}
The CM with vacuum environmental state can be used to simulate the pure damping process of a single-mode field. The damping process of a single-mode field can be described by the Lindblad-type master equation, in the weak-coupling limit,
\begin{equation}
\frac{d\hat{\rho}(t)}{dt}=g\gamma(t)\left[\opa\hat{\rho} \opa^\dagger - \frac{1}{2}\{\opa^\dagger \opa,\hat{\rho}\}\right],\label{dampME}
\end{equation}
where $\opa$ is the annihilation operator, $\hat{\rho}(t)$ is the density operator of the field, $g\ll1$ is the coupling strength and $\gamma(t)$ is the damping rate. We note that the evolution of a generic Gaussian state, governed by Eq. (\ref{dampME}), can be described in terms of the covariance matrix, as shown in Eq. (\ref{covmatrix}), with the matrices $X(t)=\exp{[-\Gamma(t)/2]}I_2$ and $Y(t)=\{1-\exp{[-\Gamma(t)}]\}I_2/2$ where $\Gamma(t)=2g\int_0^t{\gamma(s)ds}$. The matrices $X(t)$ and $Y(t)$ coincide with $X_l$ and $Y_l$ in Eqs. (\ref{covmatrixX}) and (\ref{covmatrixY}), for the vacuum environment, through
\begin{equation}
\Gamma(t)=-2\log{|c_{1,1}(l)|}.
\label{Gt}
\end{equation}
Above, we have set the elapsed time $t=l\tau$ where $\tau$ is the time interval between two successive system-environment collisions as mentioned before.

The non-Markovianity of the damping master equation, ${\cal N}_{\text PD}$, is measured, basing on the indivisibility of the dynamical map, through the time-dependent damping rate $\gamma(t)$ \cite{torre2015,hall2014} as
\begin{equation}
{\cal N}_\text{PD}=-g\int_{{\cal I}'}{\gamma(t)dt},
\label{NPD}
\end{equation}
where ${\cal I}'$ are the intervals in which $\gamma(t)<0$. It has been shown that ${\cal N}_{\text{PD}}$ is proportional to the degree proposed by Rivas {\it et al}. which is measured by the increases in entanglement \cite{rivas2010}. Eq. (\ref{NPD}) indicates that the nonzero non-Markovianity originates from the negative $\gamma(t)$ during the evolution. The correspondence between the damping rate in Eq. (\ref{dampME}) and $c_{1,1}(l)$ in the stroboscopic CM is obtained as, via Eq. (\ref{Gt}),
\begin{equation}
\gamma(t)=\frac{d\Gamma(t)}{dt}\sim -\log{\Big|\frac{c_{1,1}(l)}{c_{1,1}(l-1)}\Big|}.
\end{equation}
Apparently, the necessary and sufficient condition of the non-Markovianity of the pure damping process, i.e. $\gamma(t)<0$, is consistent with ours in the CM, i.e. $|c_{1,1}(l)|>|c_{1,1}(l-1)|$.

The contribution of the negative damping rate to the non-Markovian dynamics can be interpreted by the reverse quantum jump in the theory of non-Markovian quantum jump \cite{piilo2008, addis2014}. A quantum jump, occurring at positive $\gamma(t)$, always interrupts the deterministic evolution, while the reverse jump, occurring at negative $\gamma(t)$, will recover the coherence of the system of interest. In our CM, there is a similar process to the reverse jump in the non-Markovian evolution. Remind that the physical interpretation of $|c_{1,1}(l)|$ is the contribution of the input system mode to the output of the system mode. In the Markovian evolution, $|c_{1,1}(l)|$ decreases monotonically since the photons are always leaking. Contrastively, the nonmonotonic behavior of $|c_{1,1}(l)|$ means a photon reabsorption at some intermediate steps reminiscing the reverse jump.

\subsection{Generic Gaussian environmental state}\label{sec3C}
We now consider the case that the environmental state is a generic Gaussian state. By substituting Eq. (\ref{GaussianABC}) into Eq. (\ref{lambda12}), we obtain the non-Markovianity, ${\cal N}_{\text G}(L)$, as the following,
\begin{equation}
{\cal N}_{\text G}(L) = (2n_E+1)\cosh{(2r_E)}{\cal N}_{\text{vac}}(L),\label{NG}
\end{equation}
where $n_E$ is the thermal photon number and $r_E$ is the squeezing strength of the environmental states. One can see a generic Gaussian environmental states will enhance the non-Markovianity of vacuum environment and will not modify the boundary between Markovian and non-Markovian regions.

\subsection{Entangled environmental state}\label{sec3D}
In this subsection we will investigate effects of the entanglement embedded in the block on the non-Markovianity. We restrict our investigation to the case of $L_B=2$. The entanglement of the two-mode Gaussian state can be well characterized with the logarithmic negativity \cite{vidal2002}, which measures the entanglement by quantifying the violation of positive partial transpose separability criterion and has been proved to be a full entanglement monotone \cite{plenio2005}.

Let us consider that the two environment modes in the block are in a two-mode squeezed vacuum (TMSV) state with squeezing parameter $\xi$. A TMSV state $|\text{TMSV}(\xi)\rangle_l$ is generated from the vacuum via a two-mode squeezing operator,
\begin{equation}
|\text{TMSV}(\xi)\rangle_l=\exp{\left(\frac{1}{2}\xi^*\opa_{l,1}\opa_{l,2}-\frac{1}{2}\xi\opa_{l,1}^\dagger\opa_{l,2}^\dagger \right)}|\text{vac}\rangle_l.
\label{TMSVs}
\end{equation}
where the subscript $l$ denotes the $l$-th block and $|\text{vac}\rangle_l$ stands for the vacuum state. Without loss of generality, we set $\xi$ to be real, the characteristic function of Eq. (\ref{TMSVs}) can be expressed as
\begin{eqnarray}
\chi^{\text{in}}_l(\nu_{l,1},\nu_{l,2})&=&\exp{\left(-\frac{|\nu_{l,1}|^2+|\nu_{l,2}|^2}{2}\cosh{\xi}\right)}\cr\cr
&&\times\exp{\left(\frac{\nu_{l,1}\nu_{l,2}+\nu_{l,1}^*\nu_{l,2}^*}{2}\sinh{\xi}\right)}.
\label{TMSVch}
\end{eqnarray}
The entanglement of the TMSV state measured by the logarithmic negativity is $2\xi$ \cite{ferraro2005}.

Substituting Eq. (\ref{TMSVch}) into Eq. (\ref{chiin}) and following the procedures of computing $\Lambda_l$, we can obtain the eigenvalues of $\Lambda_l$ as,
\begin{eqnarray}
\lambda_{l,\pm}&=&-\frac{1}{2}\frac{1}{c_{1,1}^2(l-1)}\left[\cosh{(2\xi)}\left(c_{1,1}^2(l)-c_{1,1}^2(l-1)\right)\right. \cr\cr
&&\pm \left.\sqrt{4|\gamma(l)-\gamma(l-1)|^2+|c_{1,1}^2(l)-c_{1,1}^2(l-1)|^2}\right],
\end{eqnarray}
where $\gamma(l-1)=\sinh{(2\xi)}c^2_{1,1}(l-1)\sum_{l'=1}^{l}{\left[c_{1,2l'}(l)c_{1,2l'+1}(l)\right]}$ and $\gamma(l)=\sinh{(2\xi)}c^2_{1,1}(l)\sum_{l'=1}^{l-1}{\left[c_{1,2l'}(l-1)c_{1,2l'+1}(l-1)\right]}$.
\begin{figure}
  \includegraphics[width=1\linewidth]{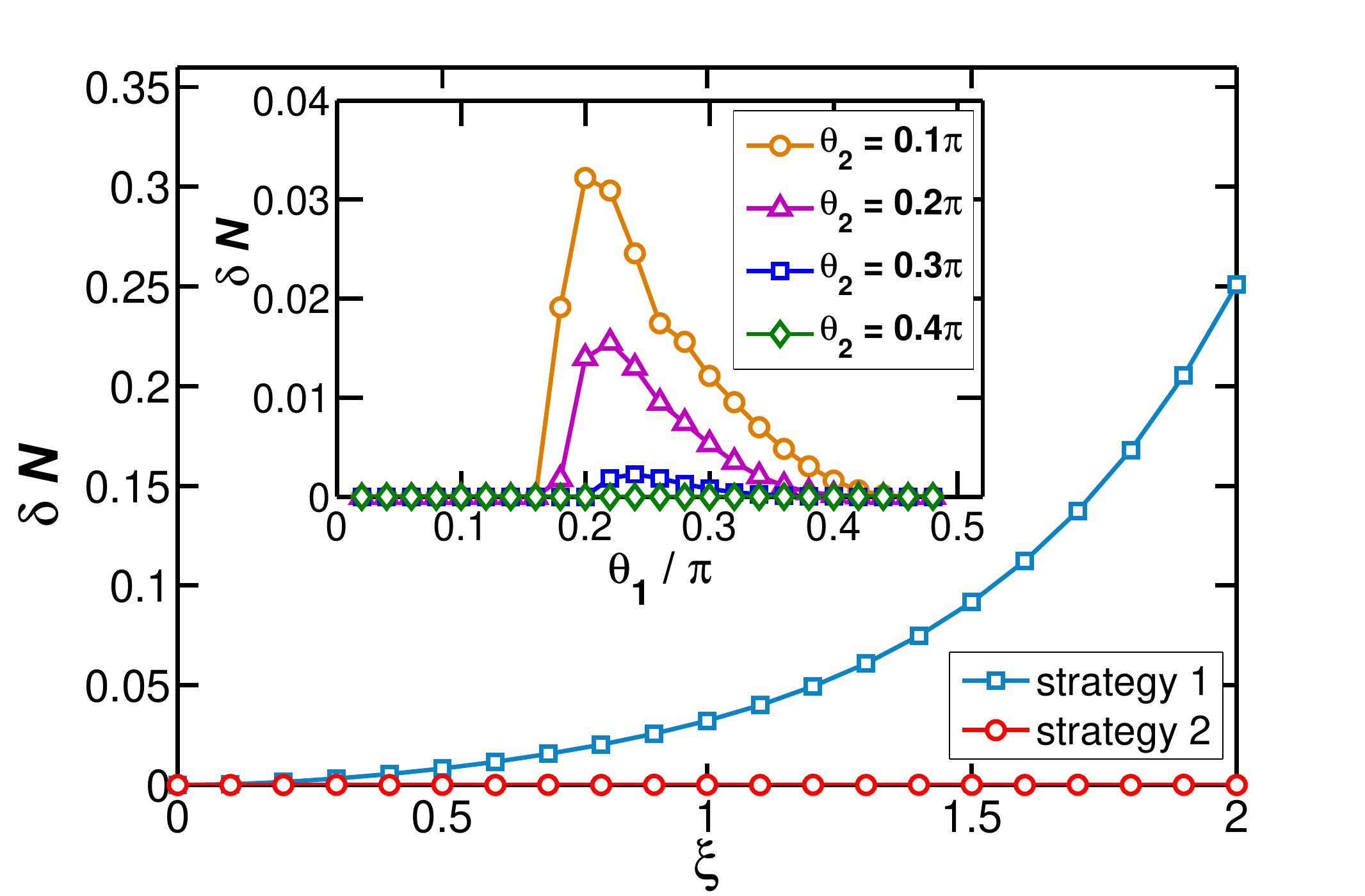}
  \caption{Non-Markovianity of the channel with environmental block in the TMSV state. The squeezing strength of the two-mode squeezing operator is $\xi = 1$. The parameters of the BSs are $\theta_1/\pi=0.2$ and $\theta_2/\pi=0.1$. For different parameters we have chosen various $L$ ensure the non-Markovianities converged.}
  \label{dNM1_1}
\end{figure}

The reduced state of each mode in a TMSV state is a thermal state $\rho_{\text{th}}$ with an effective particle number $n_{\text{th}}=\sinh^2{(\xi)}$. In order to investigate the effect of the entanglement embedded in the block, we compare the non-Markovianities of the channel with the states of the $l$-th block being the entangled state $\rho_{l}=|\text{TMSV}(\xi)\rangle_l\langle\text{TMSV}(\xi)|$ and the product state $\rho_l=\rho_{\text{th}}\otimes\rho_{\text{th}}$. We denote the non-Markovianities of each case as ${\cal N}_{\text{TMSV}}$ and ${\cal N}_{\text{prod}}$, respectively, and the discrepancy $\delta\cal{N}={\cal N}_{\text{TMSV}}-{\cal N}_{\text{prod}}$.

In Fig. \ref{dNM1_1}, we show $\delta {\cal N}$ for both strategies as functions of $\xi$ with $\theta_1/\pi=0.2$ and $\theta_2/\pi=0.1$. For strategy 1, the non-Markovianity increases with $\xi$ increasing. This indicates that the entanglement of the environmental particles in a block may enhance the non-Markovianity with the chosen parameters. In the inset of Fig. \ref{dNM1_1}, we show $\delta {\cal N}$ as functions of $\theta_1$ for different $\theta_2$ with $\xi = 1$. Although the entanglement enhances, even maximally at some optimal $\theta_1$, the non-Markovianity for $\theta_2/\pi=0.1$, $0.2$ and $0.3$, it does not affect the degree of non-Markovianity for $\theta_2/\pi=0.4$. Whether the entanglement will affect the non-Markovianity depends on the intrinsic properties of the BSs. For strategy 2, the value of $\delta {\cal N}$ is always zero and irrelevant to $\xi$ indicating that the entanglement of environment particle does not affect the degree of non-Markovianity.

\section{Conclusions}
We have presented an extended CM to simulate the non-Markovian dynamics of a quantum system. In such a CM, the unit to represent the environment is a block consisted of a number of particles. The introduced environmental block enables us to study the non-Markovianity of a quantum channel through different strategies of the system-environmental interactions and states of the environmental units.

In our CM, the system-environment ($S$-$B_l$) collisions are implemented in two strategies: in strategy 1, the system mode $S$ sequentially interacts with the environmental modes $E_{l,j}$ in the ascending order of $j$ for all $l$; in strategy 2, the system mode interacts with the environmental modes $E_{l,j}$ in the ascending order of $j$ for odd $l$ and in the descending order of $j$ for even $l$. We have adopted an all-optical system to implement the modified CM. By restricting the input modes to be Gaussian and the interactions to be linear, the dynamics of the system can be described via a Gaussian channel. With the help of the measure of non-Markovianity based on the indivisibility of dynamical maps, we have studied the effects of both strategies on the dynamics of the system mode. In strategy 1, it is shown that the non-Markoviantiy will be suppressed and converge with the size of block increasing. While in strategy 2, the non-Markovianity is independent on the size of the block.

We have also presented a necessary and sufficient condition of the non-Markovianity of the Gaussian channel. The physics behind the condition is that the contribution of the input system mode to the output of the system is nonmonotonic during the stroboscopic evolution, i.e. $|c_{1,1}(l)|>|c_{1,1}(l-1)|$ for some intermediate $l$. Such a process is similar to the reverse jump in the theory of non-Markovian quantum jump. Our measure of non-Markovianity is based on quantifying the extent by which the intermediate process fails to be CP. This corresponds to the quantification of the negative eigenvalues of the symmetric matrix associated with the intermediate process $\Phi_{l,l-1}$. This measure coincides with other existing criterions, e.g. the one based on the quantifying the negative decoherence rate of the master equation in the canonical form \cite{hall2014}, in detecting the non-Markovian features. However, since based on different point of views the existing measures may not agree with each other in quantifying the non-Markovianity of some specific channels, for instance the Gaussian channel with thermal environment \cite{strasberg2017a}. It would be interesting to investigate the connections of our measure to other ones in the future work.

We have found that the generic Gaussian environment states with nonzero temperature and squeezing will quantitatively enhance the non-Markovianity of the channel with vacuum state. We have also investigated the effects of the entanglement embedded in the environmental block on the non-Markovianity. By comparing non-Markovianity in the cases of the environmental block being in TMSV state and the product state of the corresponding reduced (thermal) states, we found that, in strategy 1, if the entanglement will enhance the non-Markovianity depends on the intrinsic properties of the channel, i.e., the reflectivity and transmissivity of the BSs. However, in strategy 2, the entanglement does not play roles in the non-Markovianity.

We emphasize that, the environment, which is in permanent contact with the system in a realistic process, is modeled by an ensemble of identical blocks in the CM. Thus we can simulate various dynamics of the open system subjected to different reservoirs by specifying the states of system and environmental blocks. For instance, apart from the Gaussian channel, we can set the environment to be vacuum and at most one excitation in the system mode to simulate the qubit amplitude-damping channel \cite{nielsen2000}.

Finally we would like to briefly discuss the possible experimental realization of our model. It could be implemented in the advanced integrated photonic quantum simulator \cite{politi2008,aspuruguzik2012,crespi2011,crespi2012,tillmann2013}. Such a platform has the advantages of intrinsic phase stability, arbitrary control of the reflectivity (and transmissivity), and flexible scalability. The integrated photonic simulator has been used to observe the Anderson localization in disordered quantum walk composed of eight steps \cite{crepsi2013}. Such a scale of the concatenated interferometers is capable to witness the effects of the interaction strategies and entanglement on the non-Markovianities with $L_B = 2$ and $L = 4$ in our model.

\acknowledgments
J. J. acknowledges supports from the National Natural Science Foundation of China No. 11747317, No. 11605022 and No. 11547119, Natural Science Foundation of Liaoning Province No. 2015020110, the Xinghai Scholar Cultivation Plan and the Fundamental Research Funds for the Central Universities, C. s. Y. from the National Natural Science Foundation of China No. 11747317, No.11775040 and No. 11375036, and the Xinghai Scholar Cultivation Plan.

\begin{appendix}
\section{Derivation of Eq. (\ref{chiout})}
Here we show the input-output relation of the joint characteristic function after $L$ times system-environment collisions. The channel composed of an array of beam-splitters maps the input modes $\hat{\textbf{a}}^{\text{in}}:=[\opa_S^{\text{in}},\opa_{1,1}^{\text{in}},...,\opa_{l,j}^{\text{in}},...,\opa_{L,L_B}^{\text{in}}]^{\text{T}}$ into the output modes $\hat{\textbf{a}}^{\text{out}}:=[\opa_S^{\text{out}},\opa_{1,1}^{\text{out}},...,\opa_{l,j}^{\text{out}},...,\opa_{L,L_B}^{\text{out}}]^{\text{T}}$ by the following transformation
\begin{equation}
\hat{\textbf{a}}^{\text{out}} = {\cal S}(L)\hat{\textbf{a}}^{\text{in}}.
\label{bogotrans}
\end{equation}
 ${\cal S}(L):=\{c_{i,j}\}$ is the ($L_BL+1$)-dimensional scattering matrix as defined in Eq. (\ref{ScatteringMatrix}) with $c_{i,j}$ ($i,j=1,2,...,L_BL+1$) being the element located at the $i$-th row and the $j$-th column. As clarified in the main text, $i = 1$ denotes the system mode and $i=2,3,...,L_BL+1$ denote the ($1,1$)-, ($1,2$)- ,...,($L,L_B$)-th environmental modes, respectively.

Recall that the channel maps, in the Schr\"{o}dinger picture, the joint input state $\hat{\rho}^{\text{in}}_J$ to the output joint state as $\hat{\rho}^{\text{out}}_J=\hat{U}\hat{\rho}^{\text{in}}_J\hat{U}^\dagger$ and, in the Heisenberg picture, the $i$-th input mode operator $\hat{a}_i^{\text{in}}$ to the output mode $\hat{a}_i^{\text{out}}=\hat{U}^\dagger\hat{a}_i^{\text{in}}\hat{U}$ with the help of the unitary evolution operator $\hat{U}$. Moreover, considering Eq. (\ref{bogotrans}), we have
\begin{equation}
\hat{U}^\dagger\hat{a}_i^{\text{in}}\hat{U} = \sum_{j=1}^{L_BL+1}{c_{i,j}\hat{a}^{\text{in}}_j}.
\end{equation}

The output joint characteristic function is calculated by
\begin{eqnarray}
\chi_J^{\text{out,L}}(\vec{\nu})&=&\text{tr}\left[\hat{\rho}_J^{\text{out}}\bigotimes_{i=1}^{L_BL+1}{\hat{D}_{\opa_i^{\text{in}}}(\nu_i)}\right]\cr\cr
&=&\text{tr}\left[\hat{U}\hat{\rho}_J^{\text{in}}\hat{U}^\dagger\bigotimes_{i=1}^{L_BL+1}{\hat{D}_{\opa_i^{\text{in}}}(\nu_i)}\right]\cr\cr
&=&\text{tr}\left[\hat{\rho}_J^{\text{in}}\bigotimes_{i=1}^{L_BL+1}{\hat{U}^\dagger\hat{D}_{\opa_i^{\text{in}}}(\nu_i)\hat{U}}\right]\cr\cr
&=&\text{tr}\left\{\hat{\rho}_J^{\text{in}}\exp{\left[\sum_{i=1}^{L_BL+1}{\sum_{j=1}^{L_BL+1}{c^*_{j,i}\nu_i\left(\opa_i^{\text{in}}\right)^\dagger-\text{h.c.} }}\right]}\right\}\cr\cr
&=&\text{tr}\left[\hat{\rho}_J^{\text{in}}\bigotimes_{i=1}^{L_BL+1}{\hat{D}_{\opa_i^{\text{in}}}\left(\sum_{j=1}^{L_BL+1}{c^*_{j,i}\nu_i} \right)} \right]\cr\cr
&=&\chi_J^{\text{in}}\left[{\cal S}^{-1}(L)\vec{\nu}\right],
\end{eqnarray}
where $\hat{D}_{\opa_i^{\text{in}}}(\nu_i) = \exp{\left[\nu_i\left(\opa_i^{\text{in}}\right)^\dagger - \nu_i^*\opa_i^{\text{in}}\right]}$ and we have used the fact ${\cal S}^{\dagger}(L)={\cal S}^{-1}(L)$.  Note that ${\cal S}(L)$ is real we obtain Eq. (\ref{chiout}).

\end{appendix}

\end{document}